\documentclass[titlepage]{amsart}
\usepackage{breakurl}             

\usepackage{amsmath,amssymb,amsfonts,amsthm,latexsym,stmaryrd,bm}
\usepackage{graphicx}
\usepackage{tikz,ifdraft,diagrams}

\let\ifmpfig\iffalse

\newtheorem{theorem}{Theorem}[section]

\theoremstyle{definition}
\newtheorem{definition}[theorem]{Definition}

\theoremstyle{remark}
\newtheorem{example}[theorem]{Example}

\newcommand*{\set}[2]{\{\,{\textstyle #1}\mathrel|{\textstyle #2}\,\}}
\newcommand*{\fin}{\mathrm{fin}}
\renewcommand*{\det}{\mathrm{det}}
\newcommand*{\ndet}{\mathrm{ndet}}
\newcommand*{\prob}{\mathrm{prob}}
\newcommand*{\id}{\mathrm{id}}
\newcommand*{\PSet}{\mathcal P}

\newcommand*{\bbC}{\mathbb C}
\newcommand*{\bbD}{\mathbb D}

\newcommand*{\bbF}{\mathbb F}
\newcommand*{\bbG}{\mathbb G}

\newcommand*{\bbN}{\mathbb N}

\newcommand*{\bbR}{\mathbb R}

\newcommand*{\obj}{\mathrm{obj}}
\newcommand*{\arr}{\mathrm{arr}}

\title{A Compositional Coalgebraic Semantics for Strategic Games\footnote{We thank 
David Colander, Samson Abramsky, Alexander Kurz, Dusko Pavlovic, Paulo Oliva and Jules Hedges for discussions
as well as the participants at the Games for Logic and Programming Languages VIII Workshop 2013 in London,
the 19th International Conference on Computing in Economics and Finance in Vancouver 2013 and the Dagstuhl Seminar Coalgebraic Semantics of Reflexive Economics, 2015.}}
\author{Achim Blumensath}
\address{Faculty of Informatics, Masaryk University, Czech Republic}
\email{blumens@fu.muni.cz}
\author{Viktor Winschel}
\address{OICOS GmbH, Mannheim}
\email{viktor.winschel@gmail.com}

\begin{document}

\begin{abstract}
We provide a compositional coalgebraic semantics for strategic games.
In our framework, like in the semantics of functional programming languages,
coalgebras represent the observable behaviour of systems derived from the behaviour of the parts over an unobservable state space.
We use coalgebras to describe and program stage games, finitely and potentially infinitely repeated hierarchical or parallel games with imperfect and incomplete information 
based on deterministic, non-deterministic or probabilistic decisions of learning agents in possibly endogenous networks.
Our framework is compositional in that arbitrarily complex network of games can be composed.
The coalgebraic approach allows to represent self-referential or reflexive structures 
like institutional dynamics, strategic network formation from within the network, belief formation, learning agents
or other self-referential phenomena that characterise complex social systems of cognitive agents.
And finally our games represent directly runnable code in functional programming languages
that can also be analysed by sophisticated verification and logical tools of software engineering.
\end{abstract}
\maketitle

\section{Introduction}
We provide a coalgebraic semantic for strategic games based on categorical methods
\cite{mac_lane_categories_1998, abramsky_introduction_2011,rutten_universal_2000,jacobs_introduction_2012}.
The first formulation of games in terms of coalgebraic semantics appeared in \cite{pavlovic_semantical_2009},
later work comprise \cite{honsell_conway_2011,lescanne_backward_2012,abramsky_coalgebraic_2012}.
In addition to these approaches we provide operations in order to compose arbitrarily complicated games from more basic ones.
We allow the players to decide based on deterministic, non-deterministic or probabilistic algorithms given their epistemic state that arises from observations,
i.e. we feature econometric or learning agents.

The coalgebraic approach to semantics has evolved for programming languages  that are modelled as abstract unobservable state transition 
systems \cite{schmidt_denotational_1986,plotkin_structural_2004,rutten_universal_2000,jacobs_introduction_2012}.
Being build on this mathematical framework our games are directly implementable for example in Haskell.
The key idea that we want to exploit in this paper is an analogy to bialgebraic theories 
\cite{klin_bialgebras_2011} covering the behaviour of programming languages.
The first usage of bialgebras for a semantic of cellular automatons for multi-agent systems was developed in \cite{trancon_y_widemann_distributive-law_2011}. We extend these ideas to game theory with the operations in the cellular automatons being players
and the cellular grid being a network of players. The semantics of cellular automatons is here then a semantics of the overall
game played by the coalgebraic players.

The coalgebraic constructions allow not only for infinite horizons or repetitions of games but also for infinite reflexive structures like beliefs of beliefs and so on that arise in economic game theory
as Harsanyi type spaces \cite{harsanyi_games_1967,harsanyi_games_1968,harsanyi_games_1968-1,mertens_formulation_1985, heifetz_topology-free_1998}.
This structure has been formulated coalgebraically in computer science \cite{moss_harsanyi_2004, moss_final_2006}.
Reflexive structures may also arise as games within networks that are played over the very structure of the network itself.
Reflexivity naturally arise in systems of cognitive agent who reason about the system they are part of, 
see for a socialogical account thereof in \cite{luhmann_gesellschaft_1998}.
The resulting mathematical paradoxes are discussed 
in \cite{lawvere_diagonal_1969,yanofsky_universal_2003,abramsky_lawvere_2010}.

Finally, the coalgebraic approach interfaces to the program verification tools or specification languages like modal logics \cite{kurz_coalgebras_2006}
that can be used to analyse the strategic games in our framework.


\section{Framework}   

In order to keep the category theoretical overhead to a minimum,
we will introduce all category theoretical notions only in the special case
of the category of sets where the objects are sets and the arrows are total functions.
Other examples of categories contain sets and relations,
measurable spaces and measurable functions,
or topological spaces and continuous functions.

The main concepts we will need
are those of a category, functor and natural transformation.

\begin{definition}
A \emph{category}~$\mathcal C$ consists of a class $\mathcal C^\obj$
of \emph{objects}, a class $\mathcal C^\arr$ of \emph{arrows,}
and a \emph{composition operation}~$\circ$ on arrows.
Each arrow $f \in \mathcal C^\arr$ has a \emph{domain} $X \in \mathcal C^\obj$
and a \emph{codomain} $Y \in \mathcal C^\obj$.
We write $f : X \to Y$ to indicate that $f$~is an arrow with domain~$X$
and codomain~$Y$.
The composition operation is assumed to satisfy the following two conditions:
\begin{enumerate}
\item The composition $f \circ g$ of two arrows is defined if, and only if,
  the domain of~$f$ is equal to the codomain of~$g$.
\item The composition operation is associative, i.e., for all arrows
  $f : X \to Y$, $g : Y \to Z$, $h : Z \to W$, 
  \begin{align*}
    (h \circ g) \circ f = h \circ (g \circ f)\,.
  \end{align*}
\item For each object $X$, there is an \emph{identity arrow} $\id_X : X \to X$
  such that
  \begin{align*}
    f \circ \id_X = f
    \quad\text{and}\quad
    \id_X \circ g = g\,,
  \end{align*}
  for all arrows $f : X \to Y$ and $g : Z \to X$.
\end{enumerate}
\end{definition}

\begin{definition}
A \emph{functor} $\bbF$ (from the category of sets to itself)
is an operation assigning
\begin{itemize}
\item to each set~$X$ a new set $\bbF(X)$ and
\item to each function $g : X \to Y$ a function $\bbF(g) : \bbF(X) \to \bbF(Y)$
\end{itemize}
such that
\begin{align*}
  \bbF(\id_X) = \id_{\bbF(X)}
  \quad\text{and}\quad
  \bbF(f \circ g) = \bbF(f) \circ \bbF(g)\,,
\end{align*}
for all sets~$X$ and all functions $f : Y \to Z$ and $g : X \to Y$.
\begin{center}
\ifmpfig
\begin{mpfig}
  u := 1.6cm;

  z0 = (0,u);
  z1 = (0,0);
  z2 = (u,0);
  z3 = (2u,u);
  z4 = (2u,0);
  z5 = (3u,0);

  pickup pencircle scaled 0.6pt;

  drawarrow 1/4[z0,z1] -- 3/4[z0,z1];
  drawarrow 1/4[z0,z2] -- 3/4[z0,z2];
  drawarrow 1/4[z1,z2] -- 3/4[z1,z2];
  drawarrow 1/4[z3,z4] -- 3/4[z3,z4];
  drawarrow 1/4[z3,z5] -- 3/4[z3,z5];
  drawarrow 1/4[z4,z5] -- 3/4[z4,z5];

  label (btex $X$ etex, z0);
  label (btex $Y$ etex, z1);
  label (btex $Z$ etex, z2);
  label (btex $\bbF(X)$ etex, z3);
  label (btex $\bbF(Y)$ etex, z4);
  label (btex $\bbF(Z)$ etex, z5);

  label lft (btex $g$               etex, 1/2[z0,z1]);
  label urt (btex $f \circ g$       etex, 1/2[z0,z2]);
  label bot (btex $f$               etex, 1/2[z1,z2]);
  label lft (btex $\bbF(g)$         etex, 1/2[z3,z4]);
  label urt (btex $\bbF(f \circ g)$ etex, 1/2[z3,z5]);
  label bot (btex $\bbF(f)$         etex, 1/2[z4,z5]);
\end{mpfig}
\else
\includegraphics{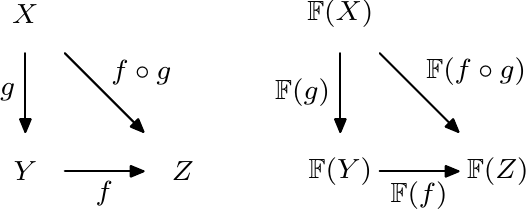}
\fi
\end{center}
\end{definition}
As an example, let us introduce three functors that will be used below.
\begin{enumerate}
\item The \emph{identity functor}~$\id$ maps every set~$X$ and every
  function $f : X \to Y$ to itself.
\item The \emph{finite power-set functor} $\PSet_\fin$ maps
  every set~$X$ to the set $\PSet_\fin(X)$ of its finite subsets,
  and it maps a function $f : X \to Y$ to the function
  \begin{align*}
    \PSet_\fin(f) : \PSet_\fin(X) \to \PSet_\fin(Y)
                  : S \mapsto \set{ f(s) }{ s \in S }\,.
  \end{align*}
\item The \emph{finite probability functor} $\bbD_\fin$
  maps a set~$X$ to the set of all finite probability distributions on~$X$,
  i.e., all maps $d : X \to [0,1]$ such that only finitely many
  elements of~$X$ are mapped to non-zero values.
  For a function $f : X \to Y$, it returns the function
  \begin{align*}
    \bbD_\fin(f) : \bbD_\fin(X) \to \bbD_\fin(Y)
                 : d \mapsto d_f\,,
  \end{align*}
  where
  \begin{align*}
    d_f(y) := \sum_{x \in f^{-1}(y)} d(x)\,.
  \end{align*}
\end{enumerate}

Beside the notion of a functor, we also need those of a natural transformation
and a distributive law.
\begin{definition}
(a)
A \emph{natural transformation} $\eta : \bbF \Rightarrow \bbG$ 
from a functor~$\bbF$ to a functor~$\bbG$ is a family $\eta = (\eta_X)_X$
of functions
\begin{align*}
  \eta_X : \bbF(X) \to \bbG(X)\,,
\end{align*}
indexed by sets~$X$, satisfying
\begin{align*}
  \eta_Y \circ \bbF(f) = \bbG(f) \circ \eta_X\,,
  \quad\text{for every function } f : X \to Y\,.
\end{align*}
\begin{center}
\ifmpfig
\begin{mpfig}
  u := 1.5cm;

  z0 = (0,u);
  z1 = (0,0);
  z2 = (u,u);
  z3 = (u,0);
  z4 = (2u,u);
  z5 = (2u,0);

  pickup pencircle scaled 0.6pt;

  drawarrow 1/4[z0,z1] -- 3/4[z0,z1];
  drawarrow 1/4[z2,z3] -- 3/4[z2,z3];
  drawarrow 1/4[z2,z4] -- 3/4[z2,z4];
  drawarrow 1/4[z3,z5] -- 3/4[z3,z5];
  drawarrow 1/4[z4,z5] -- 3/4[z4,z5];

  label (btex $X$       etex, z0);
  label (btex $Y$       etex, z1);
  label (btex $\bbF(X)$ etex, z2);
  label (btex $\bbF(Y)$ etex, z3);
  label (btex $\bbG(X)$ etex, z4);
  label (btex $\bbG(Y)$ etex, z5);

  label lft (btex $f$       etex, 1/2[z0,z1]);
  label lft (btex $\bbF(f)$ etex, 1/2[z2,z3]);
  label rt  (btex $\bbG(f)$ etex, 1/2[z4,z5]);
  label top (btex $\eta_X$  etex, 1/2[z2,z4]);
  label bot (btex $\eta_Y$  etex, 1/2[z3,z5]);
\end{mpfig}
\else
\includegraphics{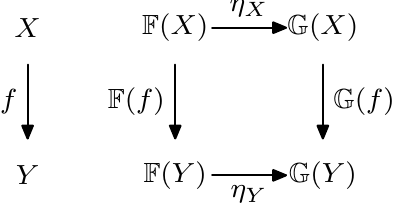}
\fi
\end{center}

(b) A \emph{distributive law} between two functors $\bbF$~and~$\bbG$
is a natural transformation
$\eta : \bbF \circ \bbG \Rightarrow \bbG \circ \bbF$.
\end{definition}

Examples of natural transformations will appear in
Section~\ref{Sect:transformations} below.

\subsection{Processes}   

Before introducing games, let us define the simpler notion of a process,
which corresponds to a game with a single player.
Processes will provide the technical machinery our framework is based on.

A process is a state based system transforming an input sequence into
an output sequence. In each step it receives an input value and,
depending on its current state, it produces an output value and changes
its state. Alternatively, a process can decide to terminate.
Formally, a \emph{process} is given by
\begin{itemize}\itemsep=0pt\parskip=0pt
\item a set $S$ of \emph{states,}
\item a set $I$ of \emph{inputs,}
\item a set $O$ of \emph{outputs,}
\item a set $R$ of \emph{results,} and
\item a function
  $\pi : S \times I \to \bbC(R + S \times O)$,
  for some functor~$\bbC$.
\end{itemize}
The function~$\pi$ describes one step of the process.
When in state $s \in S$ and given the input $i \in I$,
the process chooses a possible continuation that consists in
either terminating with a result $r \in R$,
or in continuing in a state $s' \in S$ and producing an output value $c \in O$.

In the above definition, the \emph{choice functor}~$\bbC$ determines which kind
of process we are dealing with. Important examples for choice functors are the
following ones.

\begin{enumerate}
\item The \emph{deterministic choice functor}
  $\bbC_\det= \id$
  is the identity functor. It can be used if the input
  uniquely determines what happens next.
\item The \emph{non-deterministic choice functor}
  $\bbC_\ndet = \PSet_\fin$
  is the finite power-set functor.
  It can be used if, for a given input, there might be several
  possible continuations of the process.
\item The \emph{probabilistic choice functor}
  $\bbC_\prob = \bbD_\fin$
  is the finite probability functor.
  It can be used if the continuation of the process is random.
\end{enumerate}

To apply the category theoretical machinery it will be convenient
to write the function~$\pi$ in the form
\begin{align*}
  \pi : S \to (\bbC(R + S \times O))^I.
\end{align*}
In category theoretical terms, such functions can be seen as so-called
coalgebras.
\begin{definition}
Let $\bbF$~be a functor. An \emph{$\bbF$-coalgebra} is a function
$h : X \to \bbF(X)$, for some set~$X$.
\end{definition}

Hence, a process~$\pi$ becomes a $\Pi_0$-coalgebra
$\pi : S \to \Pi_0(S)$,
where $\Pi_0$~is the \emph{process} functor
\begin{align*}
  \Pi_0(X) := \bbC(R + X \times O)^I\,.
\end{align*}

We denote by
\begin{align*}
  \Pi(S;I,O,R) := \Pi_0(S)^S
\end{align*}
the set of all processes with states~$S$, inputs~$I$, outputs~$O$, and
results~$R$.

\subsection{Transformations of choice functors}   
\label{Sect:transformations}

In this section we present several natural transformations between choice
functors that will be needed in the next section.
\begin{enumerate}
\item For two choice functors $\bbC_1$~and~$\bbC_2$, we define a natural
  transformation
  \begin{align*}
    \mu_{1,2} : \bbC_1 \circ \bbC_2 \Rightarrow \bbC_{1,2}
  \end{align*}
  that combines a choice of~$\bbC_1$ followed by a choice of~$\bbC_2$
  into a single choice with respect to a combined functor~$\bbC_{1,2}$.
\item For a choice functor~$\bbC$ and fixed sets $A,B$, we define
  a distributive law
  \begin{align*}
    \delta : A + B \times \bbC(X) \Rightarrow \bbC(A + B \times X)\,.
  \end{align*}
\item For two choice functors $\bbC_1$~and~$\bbC_2$, we define a natural
  transformation
  \begin{align*}
    \lambda_{1,2} :
      \bbC_1(X) \times \bbC_2(Y) \Rightarrow \bbC_{1,2}(X \times Y)\,.
  \end{align*}
\end{enumerate}

The definitions of all three natural transformations are the ones you would
expect from looking at the respective types. We encourage the reader to
skip the formal definitions below, which are only included for the sake of
completeness.

\smallskip
(1.)
For $\bbC_2 = \bbC_\det$, we can use $\bbC_{1,2} := \bbC_1$
and the identity function
\begin{align*}
  \mu_{1,\det} : \bbC_1(X) \to \bbC_1(X)\,.
\end{align*}
Analogously, we can define $\mu_{1,2}$ for $\bbC_1 = \bbC_\det$.
For $\bbC_1 = \bbC_2 = \bbC_\ndet$, we use $\bbC_{1,2} := \bbC_\ndet$ and
the functions
\begin{align*}
  \mu_{\ndet,\ndet} : \PSet_\fin(\PSet_\fin(X)) \to \PSet_\fin(X)
                    : \textstyle U \mapsto \bigcup_{Z \in U} Z
\end{align*}
mapping a set $U \subseteq \PSet_\fin(X)$ to its union.
For $\bbC_1 = \bbC_2 = \bbC_\prob$, we use
$\bbC_{1,2} := \bbC_\prob$ and the functions
\begin{align*}
  \mu_{\prob,\prob} : \bbD_\fin(\bbD_\fin(X)) \to \bbD_\fin(X)
\end{align*}
mapping a distribution~$d$ over $\bbD_\fin(X)$ to the distribution
\begin{align*}
  x \mapsto \sum_{d' \in \bbD(X)} d(d') \cdot d'(x)\,.
\end{align*}
The case where one of $\bbC_1$ and $\bbC_2$ equals $\bbC_\ndet$ and the
other one equals $\bbC_\prob$ is more involved. We omit the definitions.

\smallskip
(2.) We define~$\delta$ as follows.
If $\bbC = \bbC_\det$, we can use the identity map
\begin{align*}
  \delta_\det : A + B \times X \to A + B \times X\,.
\end{align*}
If $\bbC = \bbC_\ndet$, we use the map
\begin{align*}
  \delta_\ndet : A + B \times \PSet_\fin(X) \to \PSet_\fin(A + B \times X)\,,
\end{align*}
defined by
\begin{align*}
  \delta_\ndet(x) :=
    \begin{cases}
      \{x\}  &\text{for } x \in A\,, \\
      \set{ (b,u) }{ u \in U }
             &\text{for } x = (b,U) \in B \times \PSet_\fin(X)\,.
    \end{cases}
\end{align*}
If $\bbC = \bbC_\prob$, we use the map
\begin{align*}
  \delta_\prob : A + B \times \bbD_\fin(X) \to \bbD_\fin(A + B \times X)\,,
\end{align*}
defined by
\begin{align*}
  \delta_\prob(x) :=
    \begin{cases}
      d_x     &\text{for } x \in A\,, \\
      d_{b,e} &\text{for } x = (b,e) \in B \times \bbD_\fin(X)\,,
    \end{cases}
\end{align*}
where
\begin{align*}
  d_x(y) := \begin{cases}
              1  &\text{for } y = x\,, \\
              0  &\text{otherwise}\,.
            \end{cases}
  \quad\text{and}\quad
  d_{b,e}(y) := \begin{cases}
                  e(c) &\text{for } y = (b,c)\,, \\
                  0    &\text{otherwise}\,.
                \end{cases}
\end{align*}

\smallskip
(3.)
For $\bbC_1 = \bbC_\det$, we can use for
\begin{align*}
  \lambda_{\det,\det} : X \times \bbC_2(Y) \Rightarrow \bbC_2(X \times Y)
\end{align*}
the distributive law~$\delta$ from (2.) (setting $A := \emptyset$ and $B := X$).
The case where $\bbC_2 = \bbC_\det$ is handled symmetrically.
For $\bbC_1 = \bbC_2 = \bbC_\ndet$, we define
\begin{align*}
  \lambda_{\ndet,\ndet}
    : \PSet_\fin(X) \times \PSet_\fin(Y) \to \PSet_\fin(X \times Y)
    : (U,V) \mapsto U \times V\,.
\end{align*}
For $\bbC_1 = \bbC_2 = \bbC_\prob$, we define
\begin{align*}
  \lambda_{\prob,\prob}
    : \bbD_\fin(X) \times \bbD_\fin(Y) \to \bbD_\fin(X \times Y)
    : (d,d') \mapsto e_{d,d'}
\end{align*}
where
\begin{align*}
  e_{d,d'}(x,y) := d(x)\cdot d'(y)\,.
\end{align*}
Again, we omit the cases mixing $\bbC_\ndet$ and $\bbC_\prob$.

\subsection{Operations on processes}   

Before introducing games, let us present several operations intended to construct
processes from simpler ones.
We start with sums and products of processes.

(a) The \emph{sum} of two processes is a process where,
depending on the state, either the first process takes a step,
or the second one does.
We only support the case where both processes use the same choice functor.
Formally, the sum~$+$ is the operation
\begin{align*}
  {+} : \Pi(S_0;I,O_0,R_0) + \Pi(S_1;I,O_1,R_1)
    \to \Pi(S_0 + S_1;I,O_0 + O_1, R_0 + R_1)
\end{align*}
defined by
\begin{align*}
  (\pi_0 + \pi_1)(s)
    := \begin{cases}
         \pi_0(s) &\text{if } s \in S_0\,, \\
         \pi_1(s) &\text{if } s \in S_1\,.
       \end{cases}
\end{align*}

(b) The \emph{product} $\pi_1 \times \pi_2$ of two processes is a process where
both components take steps simultaneously.
We support the case where $\pi_1$~and~$\pi_2$ use different choice functors.
Suppose that $\pi_1$~uses~$\bbC_1$, while $\pi_2$~uses~$\bbC_2$.
Formally, the product~$\times$ is the operation
\begin{align*}
  {\times} &: \Pi(S_0;I_0,O_0,R_0) + \Pi(S_1;I_1,O_1,R_1) \to {}\\
      &\qquad \Pi(S_0 \times S_1;\ I_0 \times I_1,\ O_0 \times O_1,\ 
                   R_0 \times R_1 + R_0 + R_1)
\end{align*}
defined by
\begin{align*}
  (\pi_0 \times \pi_1)(s_0,s_1)(i_0,i_1)
  := (\bbC_{1,2}(f) \circ \lambda_{1,2})(\pi_0(s_0)(i_0),\pi_1(s_1)(i_1))\,,
\end{align*}
where $\lambda$ is the natural transformation from
Section~\ref{Sect:transformations} and
\begin{align*}
  f : {}&(R_0 + S_0 \times O_0) \times (R_1 + S_1 \times O_1) \to {} \\
   &\quad (R_0 \times R_1 + R_0 + R_1 + S_0 \times S_1 \times O_0 \times O_1)
\end{align*}
is the function
\begin{align*}
  f(x_0,x_1) :=
    \begin{cases}
      (x_0,x_1) &\text{if } x_0 \in R_0 \text{ and } x_1 \in R_1\,, \\
      x_0       &\text{if } x_0 \in R_0 \text{ and } x_1 \notin R_1\,, \\
      x_1       &\text{if } x_0 \notin R_0 \text{ and } x_1 \in R_1\,, \\
      (s_0,s_1,c_0,c_1)
                &\text{if } x_0 = (s_0,c_0) \text{ and } x_1 = (s_1,c_1)\,,
    \end{cases}
\end{align*}

(c) We also introduce two operations to modify the inputs and outputs.
Given a process~$\pi$ and a function~$f$, we define new processes
$\pi \triangleright f$ and $f \triangleright \pi$ as follows.

For a function $f : S \times O \to S \times O'$ and a process
$\pi \in \Pi(S;I,O,R)$, the process $\pi \triangleright f$
applies, after each step, the function~$f$ to the returned state-output pair.
Formally, we define $\pi \triangleright f \in \Pi(S;I,O',R)$ by
\begin{align*}
  (\pi \triangleright f)(s)(i) :=
    \bbC(\id + f)\bigl(\pi(s)(i)\bigr)\,.
\end{align*}

For a function $f : I' \to I$ and a process
$\pi \in \Pi(S;I,O,R)$, the process $f \triangleright \pi$
applies, before each step, the function~$f$ to the given input value.
Formally, we define $f \triangleright \pi \in \Pi(S;I',O,R)$ by
\begin{align*}
  (f \triangleright \pi)(s)(i) :=
    \pi(s)(f(i))\,.
\end{align*}

(d) Finally, we introduce two more complicated operations on processes.
The \emph{feedback operation} takes a process~$\pi$ and feeds back its output
as an additional input. That is, given a process $\pi \in \Pi(S;I \times O,O,R)$
we construct a new process $\pi^\circlearrowleft \in \Pi(S \times O; I,O,R)$
which, at each step, calls the process~$\pi$ with its current input value
and the output of the previous turn.
We define
\begin{align*}
  \pi^\circlearrowleft(s,c)(i) := \bbC(\id_R + f)(\pi(s)(i,c))\,,
\end{align*}
where
\begin{align*}
  f : S \times O \to (S \times O) \times O : (s,c) \mapsto ((s,c),c)\,.
\end{align*}

(e) The \emph{cascading operation} takes two processes $\pi$~and~$\varrho$,
runs them in parallel, and uses the outputs of the first process as
inputs of the second one.
We support the case where $\pi$~and~$\varrho$ use different choice functors.
Suppose that $\pi$~uses~$\bbC_1$, while $\varrho$~uses~$\bbC_2$.
Given $\pi \in \Pi(S;I,M,P)$ and $\varrho \in \Pi(T;M,O,R)$, we define
$\pi \triangleright \varrho \in \Pi(S \times T;I,O,P + R)$ as follows.
Let
\begin{align*}
  \varrho' &: T \times M \to \bbC_2(R + T \times O)
            : (t,m) \mapsto \varrho(t)(m)\,, \\
  \pi'     &: S \times T \times I \to \bbC_1(P + S \times T \times M)
            : (s,t,i) \mapsto \bbC_1(\id_P + f_t)\bigl(\pi(s)(i)\bigr)\,,
\end{align*}
where
\begin{align*}
  f_t : S \times M \to S \times T \times M : (s,m) \mapsto (s,t,m)\,.
\end{align*}
We set
\begin{align*}
  (\pi \triangleright \varrho)(s,t)(i) :=
    \bigl(\bbC_{1,2}(g) \circ \mu \circ \delta \circ
          \bbC_1(\id_P + \id_S \times \varrho')\bigr)
      \bigl(\pi'(s,t,i)\bigr)\,,
\end{align*}
where
$\mu$~and~$\delta$ are the natural transformations from
Section~\ref{Sect:transformations} and
\begin{align*}
  g : P + S \times (R + T \times O) \to P + R + S \times T \times O
\end{align*}
is the function
\begin{align*}
  g(x) := \begin{cases}
            x       &\text{if } x \in P\,, \\
            r       &\text{if } x = (s,r) \in S \times R\,, \\
            (s,t,c) &\text{if } x = (s,(t,c)) \in S \times T \times O\,.
          \end{cases}
\end{align*}

\subsection{Games}   

We consider games between several players that can consist of finitely
many or infinitely many rounds. The game starts in a certain state
and, in each round, every player chooses
an action to perform. These actions determine the state the game enters next.
To determine the outcome of a game, we assume that it produces an output value
with each turn and that, at the end of the game, it returns some result.
Together, the produced sequence of output values and the final result will
determine the outcome.
Formally, a game is therefore given by
\begin{itemize}\itemsep=0pt\parskip=0pt%
\item a set~$N$ of \emph{players,}
\item for each player $p \in N$, a set~$A_p$ of \emph{actions} for player~$p$,
\item a set~$S$ of \emph{states} of the game,
\item a set~$R$ of \emph{results,}
\item a set~$O$ of \emph{output values,} and
\item a function
  \begin{align*}
    \gamma : S \times \prod_{p \in N} A_p \to \bbC(R + S \times O)\,.
  \end{align*}
\end{itemize}
Thus, a game is a process where the input has the special form
$\prod_{p \in N} A_p$.
In particular, a game~$\gamma$ is a $\Gamma$-coalgebra
\begin{align*}
  \gamma : S \to \Gamma(S)\,,
\end{align*}
where $\Gamma$~is the \emph{game} functor
\begin{align*}
  \Gamma(S) := \bbC(R + S \times O)^{\prod_{p \in N} A_p}\,.
\end{align*}

\begin{example}
To formalize the Prisoner's Dilemma in our framework we use
two players $N := \{1,2\}$, each with two actions $A_p := \{c,d\}$
(`confess' and `deny').
The game needs only one state $S := \{{*}\}$, no outputs $O := \emptyset$,
and results $R := \bbR \times \bbR$.
The deterministic game function $\gamma : S \to R^{A_1 \times A_2}$ is
defined by
\begin{align*}
  \gamma({*})(a_1,a_2) :=
    \begin{cases}
      (1,1)  &\text{if } (a_1,a_2) = (c,c)\,, \\
      (2,-1) &\text{if } (a_1,a_2) = (d,c)\,, \\
      (-1,2) &\text{if } (a_1,a_2) = (c,d)\,, \\
      (0,0)  &\text{if } (a_1,a_2) = (d,d)\,.
    \end{cases}
\end{align*}
\end{example}

\begin{example}\label{ex: repeated prisoner's dilemma}
Let us also formalize the Repeated Prisoner's Dilemma.
Again, there are two players $N := \{1,2\}$ with two actions $A_p := \{c,d\}$
each.
We still have only one state $S := \{{*}\}$, but now use outputs
$O := \bbR \times \bbR$ and no results $R := \emptyset$.
The deterministic game function $\gamma : S \to (S \times O)^{A_1 \times A_2}$
is defined by
\begin{align*}
  \gamma({*})(a_1,a_2) :=
    \begin{cases}
      ({*},(1,1))  &\text{if } (a_1,a_2) = (c,c)\,, \\
      ({*},(2,-1)) &\text{if } (a_1,a_2) = (d,c)\,, \\
      ({*},(-1,2)) &\text{if } (a_1,a_2) = (c,d)\,, \\
      ({*},(0,0))  &\text{if } (a_1,a_2) = (d,d)\,.
    \end{cases}
\end{align*}
\end{example}

\begin{example}
For a more involved example, we consider a social game using
endogenous networks.
Given a group~$N$ of players, we model their social interactions
as a graph $\langle N,E\rangle$ where the edge relation~$E$ connects
two players if they are friends.
In each turn of the game, new friendships may form and old ones may end.
Thus, the graph changes in the course of the game.
We can model this game in our framework by using as set of states~$S$
the set of all possible edge relations~$E$.
Each player~$p$ has two possible actions: he can befriend another player~$q$,
or he can end an existing friendship with some player.
The game function $\gamma : S \to \Gamma(S)$ takes the current network~$E$
as an input and modifies it according to the actions of all players.
\end{example}

\subsection{Players and strategies}   

Let $\gamma : S \to \Gamma(S)$ be a game.
A \emph{strategy} for a player $p \in N$ is a function
telling him which action to choose in a given turn of the game.
The player has access to his current observations and his knowledge
of the play so far.
Thus, formally a strategy is a function
\begin{align*}
  \sigma : E_p \times B_p \to \bbC(E_p \times A_p)\,,
\end{align*}
where $E_p$~is the \emph{epistemic state} of player~$p$ and
$B_p$~is the set of possible \emph{observations.}
Again, we write~$\sigma$ as a coalgebra
\begin{align*}
  \sigma : E_p \to \bbC(E_p \times A_p)^{B_p}\,,
\end{align*}
that is, a process with inputs~$B_p$, outputs~$A_p$, and
results $R = \emptyset$.

The observations of a player depend on the current input,
the output of the previous turn, and the actions of all players
during the previous turn.
To specify what exactly player~$p$ can observe, we use a function
\begin{align*}
  \beta_p : O \times \prod_{p \in N} A_p \to B_p\,,
\end{align*}
which we assume to be a part of the description of the game.
\begin{example}
Suppose we are playing the Repeated Prisoner's Dilemma.
A~probabilistic strategy for player~$1$ would be to copy the previous
action of the other player with probability $2/3$,
and to choose the other action with probability $1/3$.
We use only one state $E_1 := \{*\}$ and the observations
$B_1 := \{c,d\}$ are the previous actions of player~$2$.
\begin{align*}
  \sigma_1 : \{{*}\} \to \bbD_\fin(\{*\} \times \{c,d\})^{\{c,d\}}
           : {*} \mapsto d
\end{align*}
where
\begin{align*}
  d(x)({*},y) := \begin{cases}
                   2/3 &\text{if } x = y\,, \\
                   1/3 &\text{if } x \neq y\,.
                 \end{cases}
\end{align*}
\end{example}

If, in a game~$\gamma$, we fix strategies $(\sigma_p)_{p \in N_0}$
for a subset $N_0 \subseteq N$ of the players, we obtain a new game
with players $N \setminus N_0$.
We denote this game by $\gamma[\sigma_p]_{p \in N_0}$.
The formal definition is as follows.
For players $p \in N \setminus N_0$ where no strategy is provided,
we introduce a non-deterministic dummy strategy that, independently
of the input, always tells the player to
play some action from~$A_p$ without restricting his choice.
This strategy uses only one state. Its formal definition is
\begin{align*}
  \sigma_p : 1 \to \PSet_\fin(1 \times A_p)^{B_p}
           : i \mapsto A_p\,.
\end{align*}
With the help of these dummy strategies, we can define the desired game as
\begin{align*}
  \gamma[\sigma_p]_{p \in N_0} :=
  \Bigl[\Bigl[f \triangleright
              \prod_{p \in N} \sigma_p\Bigr]^\circlearrowleft
          \triangleright \gamma\Bigr]^\circlearrowleft
  \in \Pi\bigl(\textstyle S \times \prod_{p \in N} E_p \times O \times \prod_{p \in N} A_p,
          \emptyset,O,R\bigr)\,,
\end{align*}
where the function
\begin{align*}
  f : O \times \prod_{p \in N} A_p \to \prod_{p \in N} B_p
    : (c,\bar a) \mapsto (\beta_p(c,\bar a))_{p \in N}\,,
\end{align*}
computes the observations of each player.
The states of this new game are tuples
\begin{align*}
  (s,\bar e,c,\bar a) \in
  S \times \prod_{p \in N} E_p \times O \times \prod_{p \in N} A_p
\end{align*}
consisting of
a state~$s$ of the old game, the epistemic states~$\bar e$ of the players,
the output~$c$ of the last turn, and the actions~$\bar a$ the players
chose last turn.

\subsection{Game trees}   

Given a game~$\gamma$ and strategies~$\sigma_p$ for each player,
we would like to compute the result of the game if each player follows
her strategy.
Besides the techniques from the previous section, we need one more
definition: that of a \emph{game tree.}
Informally, a game tree is a tree containing all possible sequences of
events allowed in the game.
The formal definition is based on the notion of a \emph{final} coalgebra.
\begin{definition}
Let $\bbF$~be a functor. An $\bbF$-coalgebra
$\omega : \Omega \to \bbF(\Omega)$ is \emph{final} if,
for every $\bbF$-coalgebra $h : X \to \bbF(X)$, there exists a unique
morphism $\varphi : X \to \Omega$ such that the diagram
\begin{center}
\ifmpfig
\begin{mpfig}
  u := 1.5cm;

  z0 = (0,0);
  z1 = (1.5u,0);
  z2 = (0,u);
  z3 = (1.5u,u);

  pickup pencircle scaled 0.6pt;

  drawarrow 1/4[z0,z1] -- 3/4[z0,z1];
  drawarrow 1/4[z2,z0] -- 3/4[z2,z0];
  drawarrow 1/4[z2,z3] -- 3/4[z2,z3];
  drawarrow 1/4[z3,z1] -- 3/4[z3,z1];

  label (btex \normalsize $X$ etex, z2);
  label (btex \normalsize $\Omega$ etex, z3);
  label (btex \normalsize $\bbF(X)$ etex, z0);
  label (btex \normalsize $\bbF(\Omega)$ etex, z1);
  label lft (btex $h$ etex, 1/2[z2,z0]);
  label top (btex $\varphi$ etex, 1/2[z2,z3]);
  label bot (btex $\bbF(\varphi)$ etex, 1/2[z0,z1]);
  label rt  (btex $\omega$ etex, 1/2[z3,z1]);
\end{mpfig}
\else
\includegraphics{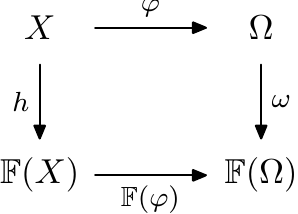}
\fi
\end{center}
commutes.
\end{definition}

For the process functors
\begin{align*}
  \Pi_0(X) = \bbC(R + X \times O)^I
\end{align*}
there exist final $\Pi_0$-coalgebras $\omega : \Omega \to \Pi_0(\Omega)$,
provided that the choice functor~$\bbC$ is sufficiently well-behaved.
In particular, this is the case
for the three choice functors $\bbC_\det$, $\bbC_\ndet$, and $\bbC_\prob$.

Let us describe the final $\Pi_0$-coalgebras for the choice
functors~$\bbC$ introduced above.
The elements of these final coalgebras are \emph{trees,}
which are directed acyclic graphs such that there exists one vertex,
the \emph{root} of the tree, with the property that every other
vertex can be reached by a unique path from the root.
A tree is \emph{$(A,B,C)$-labelled} if it has more than one vertex and
\begin{itemize}
\item the root is unlabelled,
\item every other inner vertex is labelled by an element of~$B$,
\item every leaf is labelled by an element of $A \cup B$,
\item every edge is labelled by an element of~$C$.
\end{itemize}
If there is an edge with label~$c$ from a vertex~$x$ to a vertex~$y$, we call
$y$~the \emph{$c$-successor} of~$x$.

(a)
We start with the functor
\begin{align*}
  \Pi_0(X) := (A + X \times B)^C
\end{align*}
for $\bbC = \bbC_\det$. In this case the final $\Pi_0$-coalgebra
$\omega : \Omega \to \Pi_0(\Omega)$ takes the following form.
The set $\Omega$~consists of all
$(A,B,C)$-labelled trees that are \emph{deterministic,} that is,
such that every leaf has a label in~$A$ and
every inner vertex has exactly one $c$-successor, for each $c \in C$.
The function~$\omega$ is defined as follows:
given a tree~$T$ and a value $c \in C$, we distinguish two cases
depending on the label of the $c$-successor~$x$ of the root.
If $x$~is labelled by an element $a \in A$,
we set $\omega(T)(c) := a$.
If $x$~is labelled by an element $b \in B$,
we set $\omega(T)(c) := \langle T',b\rangle$
where $T'$~is the subtree of~$T$ rooted at~$x$.

To see that this is indeed the final $\Pi_0$-coalgebra, consider
an arbitrary $\Pi_0$-coalgebra $\pi : S \to \Pi_0(S)$.
The required unique function $\varphi : S \to \Omega$ is given by
\begin{align*}
  \varphi(s) := T_s\,, \quad\text{for } s \in S\,,
\end{align*}
where the tree~$T_s$ is defined as follows:
we first construct a graph $\langle V,E\rangle$ with set of vertices
$V := A + S \times B$ and the following edges.
For every $\langle s,b\rangle \in S \times B$,
there is a $c$-labelled edge from $\langle s,b\rangle$ to $\pi(s)(c)$.
The elements of~$A$ have no outgoing edges.
The vertex labelling is the natural one: a vertex $a \in A$ gets
the label~$a$ and a vertex $\langle s,b\rangle$ gets the label $b \in B$.

The tree~$T_s$ is now obtained from the unravelling of this graph starting
at a vertex $\langle s,b\rangle$, for an arbitrary $b \in B$,
by forgetting the label~$b$ of the root.
Formally, the \emph{unravelling} of a graph $\langle V,E\rangle$
starting at a vertex~$s$ is defined as the tree consisting of all finite
paths through the graph that start at~$s$. There is an edge
between two such paths if the second one is obtained from the first one
by appending a single edge. This edge also determines the label of the edge label.
The vertex labelling of the tree is obtained by labelling each path
with the label of its end-vertex.
\begin{example}
Let $\pi \in \Pi(\{s_a,s_b\}; \{0,1\}, \{a,b\}, \emptyset)$
be the deterministic process defined by
\begin{align*}
  \pi(s_x)(y) := \begin{cases}
                   (s_a,x) &\text{if } y = 0\,, \\
                   (s_b,x) &\text{if } y = 1\,.
                 \end{cases}
\end{align*}
The (top of the) tree $\varphi(s_a)$ has the following form:
\begin{center}
\ifmpfig
\begin{mpfig}
  u := 1cm;

  z0 = (0,0);
  z1 = (-u,-u);
  z2 = (u,-u);
  z3 = (-1.5u,-2u);
  z4 = (-0.5u,-2u);
  z5 = (0.5u,-2u);
  z6 = (1.5u,-2u);

  pickup pencircle scaled 0.6pt;

  draw 1/4[z0,z1]--3/4[z0,z1];
  draw 1/4[z0,z2]--3/4[z0,z2];
  draw 1/4[z1,z3]--3/4[z1,z3];
  draw 1/4[z1,z4]--3/4[z1,z4];
  draw 1/4[z2,z5]--3/4[z2,z5];
  draw 1/4[z2,z6]--3/4[z2,z6];

  label (btex \normalsize $\bullet$ etex, z0);
  label (btex \normalsize $a$ etex, z1);
  label (btex \normalsize $a$ etex, z2);
  label (btex \normalsize $a$ etex, z3);
  label (btex \normalsize $a$ etex, z4);
  label (btex \normalsize $b$ etex, z5);
  label (btex \normalsize $b$ etex, z6);

  label ulft (btex $0$ etex, 1/2[z0,z1]);
  label urt  (btex $1$ etex, 1/2[z0,z2]);
  label ulft (btex $0$ etex, 1/2[z1,z3]);
  label urt  (btex $1$ etex, 1/2[z1,z4]);
  label ulft (btex $0$ etex, 1/2[z2,z5]);
  label urt  (btex $1$ etex, 1/2[z2,z6]);
  label (btex $\bm\vdots$ etex, z3 + 0.5u*down);
  label (btex $\bm\vdots$ etex, z4 + 0.5u*down);
  label (btex $\bm\vdots$ etex, z5 + 0.5u*down);
  label (btex $\bm\vdots$ etex, z6 + 0.5u*down);
\end{mpfig}
\else
\includegraphics{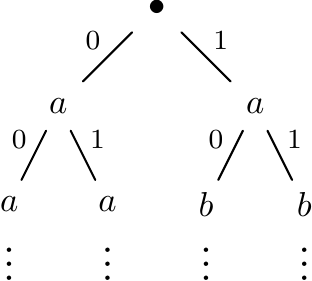}
\fi
\end{center}
\end{example}

To see that the function~$\varphi$ defined in this way has the required
property we need to check that
\begin{align*}
  \Pi_0(\varphi) \circ \pi = \omega \circ \varphi\,.
\end{align*}
For $s \in S$ and $c \in C$, suppose that
\begin{align*}
  \pi(s)(c) = \langle s',b\rangle \in S \times B\,.
\end{align*}
Let $T := \varphi(s)$ and $T' := \varphi(s')$.
Note that $T'$~is equal to the subtree of~$T$ rooted
at the $c$-successor of the root and that this $c$-successor
is labelled by~$b$.
Hence,
\begin{align*}
  (\Pi_0(\varphi) \circ \pi)(s)(c)
  = \Pi_0(\varphi)(\langle s',b\rangle)
  = \langle T',b\rangle
  = \omega(T)(c)
  = (\omega \circ \varphi)(s)(c)\,.
\end{align*}
In the case where $\pi(s)(c) = a \in A$ we argue similarly.

(b) Consider the functor
\begin{align*}
  \Pi_0(X) := \PSet_\fin(A + X \times B)^C
\end{align*}
for non-deterministic games. In this case the final $\Pi_0$-coalgebra
$\omega : \Omega \to \Pi_0(\Omega)$
takes the following form. The set $\Omega$~consists of all
$(A,B,C)$-labelled trees where each vertex has only finitely many
$c$-successors, for every $c \in C$.
The function~$\omega$ is defined as follows.
Given a tree~$T$ and a value $c \in C$,
let $S$~be the set of all $c$-successors of the root of~$T$.
Then $\omega(T)(c)$ returns the set
\begin{align*}
      \set{ a \in A }{ \text{some } x \in S \text{ has label } a }
 \cup \set{ \langle T_x,b\rangle }{ x \in S \text{ has label } b \in B }\,,
\end{align*}
where $T_x$~is the subtree of~$T$ rooted at~$x$.

Given an arbitrary $\Pi_0$-coalgebra
$\pi : S \to \Pi_0(S)$,
the required unique function $\varphi : S \to \Omega$ is defined similarly
as in~(a). We set
\begin{align*}
  \varphi(s) := T_s\,, \quad\text{for } s \in S\,,
\end{align*}
where $T_s$~is the unravelling of the following graph $\langle V,E\rangle$.
Again the set of vertices is $V := A + S \times B$ and
the vertex labelling is the natural one.
For each $\langle s,b\rangle \in S \times B$,
there is a $c$-labelled edge from~$\langle s,b\rangle$ to~$x$,
for every $x \in \pi(s)(c)$.

As above, a straightforward calculation shows that the function~$\varphi$
defined in this way has the required properties.
\begin{example}
Let $\pi \in \Pi(\{*\}; \{0,1\}, \{a,b\}, \emptyset)$
be the non-deterministic process defined by
\begin{align*}
  \pi({*})(x) := \{({*},a),({*},b)\}\,.
\end{align*}
The tree $\varphi({*})$ has the following form:
\begin{center}
\ifmpfig
\begin{mpfig}
  u := 1.4cm;

  z0  = (0,0);
  z1a = (-3u,-0.75u);
  z1b = (-1u,-0.75u);
  z2a = (1u,-0.75u);
  z2b = (3u,-0.75u);
  z3a = (-3.75u,-1.5u);
  z3b = (-3.25u,-1.5u);
  z4a = (-2.75u,-1.5u);
  z4b = (-2.25u,-1.5u);
  z5a = (-1.75u,-1.5u);
  z5b = (-1.25u,-1.5u);
  z6a = (-0.75u,-1.5u);
  z6b = (-0.25u,-1.5u);
  z7a = (0.25u,-1.5u);
  z7b = (0.75u,-1.5u);
  z8a = (1.25u,-1.5u);
  z8b = (1.75u,-1.5u);
  z9a = (2.25u,-1.5u);
  z9b = (2.75u,-1.5u);
  z10a = (3.25u,-1.5u);
  z10b = (3.75u,-1.5u);

  pickup pencircle scaled 0.6pt;

  draw 1/7[z0,z1a]--6/7[z0,z1a];
  draw 1/5[z0,z1b]--4/5[z0,z1b];
  draw 1/5[z0,z2a]--4/5[z0,z2a];
  draw 1/7[z0,z2b]--6/7[z0,z2b];
  draw 1/4[z1a,z3a]--3/4[z1a,z3a];
  draw 1/4[z1a,z3b]--3/4[z1a,z3b];
  draw 1/4[z1a,z4a]--3/4[z1a,z4a];
  draw 1/4[z1a,z4b]--3/4[z1a,z4b];
  draw 1/4[z1b,z5a]--3/4[z1b,z5a];
  draw 1/4[z1b,z5b]--3/4[z1b,z5b];
  draw 1/4[z1b,z6a]--3/4[z1b,z6a];
  draw 1/4[z1b,z6b]--3/4[z1b,z6b];
  draw 1/4[z2a,z7a]--3/4[z2a,z7a];
  draw 1/4[z2a,z7b]--3/4[z2a,z7b];
  draw 1/4[z2a,z8a]--3/4[z2a,z8a];
  draw 1/4[z2a,z8b]--3/4[z2a,z8b];
  draw 1/4[z2b,z9a]--3/4[z2b,z9a];
  draw 1/4[z2b,z9b]--3/4[z2b,z9b];
  draw 1/4[z2b,z10a]--3/4[z2b,z10a];
  draw 1/4[z2b,z10b]--3/4[z2b,z10b];

  label (btex \normalsize $\bullet$ etex, z0);
  label (btex \normalsize $a$ etex, z1a);
  label (btex \normalsize $b$ etex, z1b);
  label (btex \normalsize $a$ etex, z2a);
  label (btex \normalsize $b$ etex, z2b);
  label (btex \normalsize $a$ etex, z3a);
  label (btex \normalsize $b$ etex, z3b);
  label (btex \normalsize $a$ etex, z4a);
  label (btex \normalsize $b$ etex, z4b);
  label (btex \normalsize $a$ etex, z5a);
  label (btex \normalsize $b$ etex, z5b);
  label (btex \normalsize $a$ etex, z6a);
  label (btex \normalsize $b$ etex, z6b);
  label (btex \normalsize $a$ etex, z7a);
  label (btex \normalsize $b$ etex, z7b);
  label (btex \normalsize $a$ etex, z8a);
  label (btex \normalsize $b$ etex, z8b);
  label (btex \normalsize $a$ etex, z9a);
  label (btex \normalsize $b$ etex, z9b);
  label (btex \normalsize $a$ etex, z10a);
  label (btex \normalsize $b$ etex, z10b);

  label ulft (btex $0$ etex, 1/2[z0,z1a]);
  label lft  (btex $0$ etex, 1/2[z0,z1b]);
  label rt   (btex $1$ etex, 1/2[z0,z2a]);
  label urt  (btex $1$ etex, 1/2[z0,z2b]);
  label ulft (btex $0$ etex, 1/2[z1a,z3a]);
  label lft  (btex $0$ etex, 1/2[z1a,z3b]);
  label rt   (btex $1$ etex, 1/2[z1a,z4a]);
  label urt  (btex $1$ etex, 1/2[z1a,z4b]);
  label ulft (btex $0$ etex, 1/2[z1b,z5a]);
  label lft  (btex $0$ etex, 1/2[z1b,z5b]);
  label rt   (btex $1$ etex, 1/2[z1b,z6a]);
  label urt  (btex $1$ etex, 1/2[z1b,z6b]);
  label ulft (btex $0$ etex, 1/2[z2a,z7a]);
  label lft  (btex $0$ etex, 1/2[z2a,z7b]);
  label rt   (btex $1$ etex, 1/2[z2a,z8a]);
  label urt  (btex $1$ etex, 1/2[z2a,z8b]);
  label ulft (btex $0$ etex, 1/2[z2b,z9a]);
  label lft  (btex $0$ etex, 1/2[z2b,z9b]);
  label rt   (btex $1$ etex, 1/2[z2b,z10a]);
  label urt  (btex $1$ etex, 1/2[z2b,z10b]);
\end{mpfig}
\else
\includegraphics{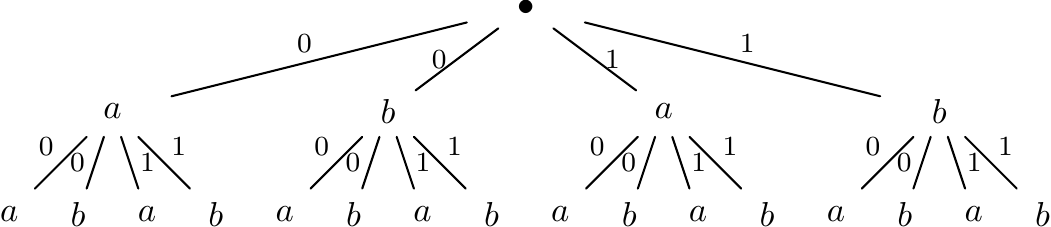}
\fi
\end{center}
\end{example}

(c) Finally, consider the functor
\begin{align*}
  \Pi_0(X) := \bbD_\fin(A + X \times B)^C
\end{align*}
for probabilistic games. In this case the final $\Pi_0$-coalgebra
$\omega : \Omega \to \Pi_0(\Omega)$
takes the following form: the set $\Omega$~consists of all
$(A,B,C \times [0,1])$-labelled trees where, for every $c \in C$
and every vertex~$v$,
\begin{itemize}
\item $v$~has only finitely many outgoing edges labelled $\langle c,p\rangle$,
  for some $p \in [0,1]$,
\item the sum of all values~$p$ such that there is an outgoing edge with label
  $\langle c,p\rangle$ equals~$1$, and
\item $v$~does not have two outgoing edges with labels
  $\langle c,p\rangle$ and $\langle c,p'\rangle$ where $p,p' \in [0,1]$
  and such that the subtrees rooted at the corresponding successors
  are isomorphic.
\end{itemize}
The function~$\omega$ is defined as follows:
given a tree~$T$, a value $c \in C$, $a \in A$, and
$\langle T',b\rangle \in \Omega \times B$, we set
\begin{align*}
  \omega(T)(c)(a) := p
\end{align*}
if the root of~$T$ has an outgoing edge with label $\langle c,p\rangle$
that leads to a leaf with label~$a$, and we set
\begin{align*}
  \omega(T)(c)(\langle T',b\rangle) := p
\end{align*}
if the root of~$T$ has an outgoing edge with label $\langle c,p\rangle$
that leads to an inner vertex~$x$ with label~$b$ such that
the subtree of~$T$ rooted at~$x$ is equal to~$T'$.
In all other cases, we set
\begin{align*}
  \omega(T)(c)(x) := 0\,.
\end{align*}

Given an arbitrary $\Pi_0$-coalgebra $\pi : S \to \Pi_0(S)$,
the required unique function $\varphi : S \to \Omega$ is defined similarly
as in~(a). We set
\begin{align*}
  \varphi(s) := T_s\,, \quad\text{for } s \in S\,,
\end{align*}
where $T_s$~is the unravelling of the following graph $\langle V,E\rangle$:
again the set of vertices is $V := A + S \times B$ and
the vertex labelling is the natural one.
For each $\langle s,b\rangle \in S \times B$ and every $x \in A + S \times B$,
there is a $\langle c,\pi(s)(c)(x)\rangle$-labelled edge
from~$\langle s,b\rangle$ to~$x$.

As above, a straightforward calculation shows that the function~$\varphi$
defined in this way has the required properties.
\begin{example}
Let $\pi \in \Pi(\{*\}; \{0,1\}, \{a,b\}, \emptyset)$
be the probabilistic process defined by
\begin{align*}
  \pi({*})(x)(s,c) := 1/2\,.
\end{align*}
The tree $\varphi({*})$ has the following form:
\begin{center}
\ifmpfig
\begin{mpfig}
  u := 1.4cm;

  z0  = (0,0);
  z1a = (-3u,-0.75u);
  z1b = (-1u,-0.75u);
  z2a = (1u,-0.75u);
  z2b = (3u,-0.75u);
  z3a = (-3.75u,-1.5u);
  z3b = (-3.25u,-1.5u);
  z4a = (-2.75u,-1.5u);
  z4b = (-2.25u,-1.5u);
  z5a = (-1.75u,-1.5u);
  z5b = (-1.25u,-1.5u);
  z6a = (-0.75u,-1.5u);
  z6b = (-0.25u,-1.5u);
  z7a = (0.25u,-1.5u);
  z7b = (0.75u,-1.5u);
  z8a = (1.25u,-1.5u);
  z8b = (1.75u,-1.5u);
  z9a = (2.25u,-1.5u);
  z9b = (2.75u,-1.5u);
  z10a = (3.25u,-1.5u);
  z10b = (3.75u,-1.5u);

  pickup pencircle scaled 0.6pt;

  draw 1/7[z0,z1a]--6/7[z0,z1a];
  draw 1/5[z0,z1b]--4/5[z0,z1b];
  draw 1/5[z0,z2a]--4/5[z0,z2a];
  draw 1/7[z0,z2b]--6/7[z0,z2b];
  draw 1/4[z1a,z3a]--3/4[z1a,z3a];
  draw 1/4[z1a,z3b]--3/4[z1a,z3b];
  draw 1/4[z1a,z4a]--3/4[z1a,z4a];
  draw 1/4[z1a,z4b]--3/4[z1a,z4b];
  draw 1/4[z1b,z5a]--3/4[z1b,z5a];
  draw 1/4[z1b,z5b]--3/4[z1b,z5b];
  draw 1/4[z1b,z6a]--3/4[z1b,z6a];
  draw 1/4[z1b,z6b]--3/4[z1b,z6b];
  draw 1/4[z2a,z7a]--3/4[z2a,z7a];
  draw 1/4[z2a,z7b]--3/4[z2a,z7b];
  draw 1/4[z2a,z8a]--3/4[z2a,z8a];
  draw 1/4[z2a,z8b]--3/4[z2a,z8b];
  draw 1/4[z2b,z9a]--3/4[z2b,z9a];
  draw 1/4[z2b,z9b]--3/4[z2b,z9b];
  draw 1/4[z2b,z10a]--3/4[z2b,z10a];
  draw 1/4[z2b,z10b]--3/4[z2b,z10b];

  label (btex \normalsize $\bullet$ etex, z0);
  label (btex \normalsize $a$ etex, z1a);
  label (btex \normalsize $b$ etex, z1b);
  label (btex \normalsize $a$ etex, z2a);
  label (btex \normalsize $b$ etex, z2b);
  label (btex \normalsize $a$ etex, z3a);
  label (btex \normalsize $b$ etex, z3b);
  label (btex \normalsize $a$ etex, z4a);
  label (btex \normalsize $b$ etex, z4b);
  label (btex \normalsize $a$ etex, z5a);
  label (btex \normalsize $b$ etex, z5b);
  label (btex \normalsize $a$ etex, z6a);
  label (btex \normalsize $b$ etex, z6b);
  label (btex \normalsize $a$ etex, z7a);
  label (btex \normalsize $b$ etex, z7b);
  label (btex \normalsize $a$ etex, z8a);
  label (btex \normalsize $b$ etex, z8b);
  label (btex \normalsize $a$ etex, z9a);
  label (btex \normalsize $b$ etex, z9b);
  label (btex \normalsize $a$ etex, z10a);
  label (btex \normalsize $b$ etex, z10b);

  label ulft (btex $0,\frac{1}{2}$ etex, 1/2[z0,z1a]);
  label rt   (btex $0,\frac{1}{2}$ etex, 1/2[z0,z1b] + 2pt*right);
  label rt   (btex $1,\frac{1}{2}$ etex, 1/2[z0,z2a]);
  label urt  (btex $1,\frac{1}{2}$ etex, 1/2[z0,z2b]);
  label ulft (btex $0,\frac{1}{2}$ etex, 1/2[z1a,z3a]);
  label urt  (btex $1,\frac{1}{2}$ etex, 1/2[z1a,z4b]);
  label ulft (btex $0,\frac{1}{2}$ etex, 1/2[z1b,z5a]);
  label urt  (btex $1,\frac{1}{2}$ etex, 1/2[z1b,z6b]);
  label ulft (btex $0,\frac{1}{2}$ etex, 1/2[z2a,z7a]);
  label urt  (btex $1,\frac{1}{2}$ etex, 1/2[z2a,z8b]);
  label ulft (btex $0,\frac{1}{2}$ etex, 1/2[z2b,z9a]);
  label urt  (btex $1,\frac{1}{2}$ etex, 1/2[z2b,z10b]);
\end{mpfig}
\else
\includegraphics{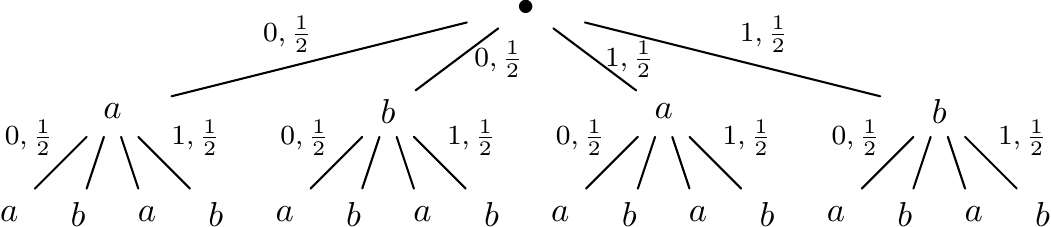}
\fi
\end{center}
(Due to space considerations we have omitted some edge labels.)
\end{example}

We have seen that the final coalgebras consist of trees
describing all possible sequences in the game.
Given a game $\gamma : S \to \Gamma(S)$ and the unique morphism
$\varphi : S \to \Omega$ into the final $\Gamma$-coalgebra,
we call the tree $\varphi(s)$ the \emph{game tree} of~$\gamma$
when starting in state~$s \in S$.

\subsection{The outcome of a game}   

After these preparations we can determine the outcome of a game.
Given a game~$\gamma$ and strategies~$\sigma_p$ for each player,
we can compute a game $\gamma[\sigma_p]_p$ without players
and determine its game tree~$T$.
Hence, it remains to define how to read off the outcome from a game tree.

Let $\gamma : S \to \Gamma(S)$ be a game without players and
let $\omega : \Omega \to \Gamma(\Omega)$ be the final $\Gamma$-coalgebra.
To define the outcome~$\gamma$ we specify a set~$U$ of \emph{outcomes}
and two functions
$\varrho : \Omega \to U$ and $\tau : \Gamma(U) \to U$ such that
\begin{center}
\ifmpfig
\begin{mpfig}
  u := 1.5cm;

  z0 = (0,0);
  z1 = (1.5u,0);
  z2 = (0,u);
  z3 = (1.5u,u);

  pickup pencircle scaled 0.6pt;

  drawarrow 1/4[z1,z0] -- 3/4[z1,z0];
  drawarrow 1/4[z2,z0] -- 3/4[z2,z0];
  drawarrow 1/4[z2,z3] -- 3/4[z2,z3];
  drawarrow 1/4[z3,z1] -- 3/4[z3,z1];

  label (btex \normalsize $\Omega$ etex, z2);
  label (btex \normalsize $U$ etex, z0);
  label (btex \normalsize $\Gamma(\Omega)$ etex, z3);
  label (btex \normalsize $\Gamma(U)$ etex, z1);
  label lft (btex $\varrho$ etex, 1/2[z2,z0]);
  label top (btex $\omega$ etex, 1/2[z2,z3]);
  label bot (btex $\tau$ etex, 1/2[z0,z1]);
  label rt  (btex $\Gamma(\varrho)$ etex, 1/2[z3,z1]);
\end{mpfig}
\else
\includegraphics{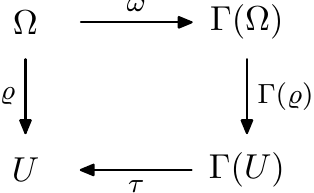}
\fi
\end{center}
Intuitively, $\varrho$~maps a game tree to its outcome, while
$\tau$~computes the outcome of a game from the outcomes of its subgames.
Hence, $\tau$~performs a local computation, while $\varrho$~is needed
to compute the limit of an infinite sequence of turns.
Ideally, the function~$\tau$ uniquely determines~$\varrho$.
This is the case, for instance, for discounted pay-off games,
where the value of a game mostly depends on an initial segment of the
game tree.

\begin{example}
Consider a deterministic two player game
with $R = \bbR \times \bbR$ and $O = \bbR \times \bbR$.
Fixing deterministic strategies for both players,
we obtain a deterministic zero-player game, the game tree of which is
either an infinite sequence over~$O$ or a finite sequence
where the last element is from~$R$ and the remaining ones are from~$O$.

Choosing a discount factor $\lambda \in (0,1)$, we can define the outcome
by the functions
\begin{align*}
  \tau : R + U \times O \to U
  \quad\text{and}\quad
  \varrho : \Omega \to U\,,
\end{align*}
where $U := \bbR \times \bbR$ and
\begin{alignat*}{-1}
  \tau(x,y) &:= (x,y)\,, &&\quad\text{for } (x,y) \in R\,, \\
  \tau((x,y),(u,v)) &:= (\lambda x + u,\lambda y + v)\,,
    &&\quad\text{for } ((x,y),(u,v)) \in U \times O\,.
\end{alignat*}
The function~$\varrho$ is uniquely determined by~$\tau$. An explicit
definition is
\begin{align*}
  \textstyle
  \varrho(x_n,y_n)_{n < \alpha} :=
    \bigl(\sum_{n < \alpha} \lambda^nx_n,
          \sum_{n < \alpha} \lambda^ny_n\bigr)\,.
\end{align*}
\end{example}

\subsection{Nash equilibria}   

Having defined the outcome of a game, we can introduce equilibria.
Consider a game~$\gamma$ with set of players~$N$ and set of outcomes
$U := \bbR^N$.
We fix an output value $\hat c \in O$ and actions $\hat a \in \prod_p A_p$
that will serve as imaginary outcome of the `first game turn'.
Let $\tau : \Gamma(U) \to U$ and $\varrho : \Omega \to U$
be the functions to compute the outcome of~$\gamma$.
For a tuple $\bar\sigma = (\sigma_p)_{p \in N}$ of strategies,
we denote by $\varphi[\bar\sigma] : S \to \Omega$ the function
from the reduced game $\gamma[\bar\sigma]$ to the final coalgebra.

Given strategies~$\sigma_p$, for each $p \in N$,
initial states $s_0 \in S$ and $e_p \in E_p$, for $p \in N$,
and a player $q \in N$,
we say that $\langle\sigma_q,e_q\rangle$ is a \emph{best response} to
$\langle\sigma_p,e_p\rangle_{p \in N \setminus \{q\}}$ in the game
$\langle\gamma,s_0\rangle$ if
\begin{align*}
  \varrho(\varphi[\bar\sigma](s_0,\bar e,\hat c,\hat a))
  \geq \varrho(\varphi[\bar\sigma'](s_0,\bar e',\hat c,\hat a))\,,
\end{align*}
for all tuples~$\bar\sigma'$ and $\bar e'$ that differ from, respectively,
$\bar\sigma$~and~$\bar e$ only in the $p$-th component.

We say that $\langle\bar\sigma,\bar e\rangle$ is a \emph{Nash equilibrium}
of $\langle\gamma,s_0\rangle$ if,
for every player $q \in N$, $\langle\sigma_q,e_q\rangle$
is a best response to
$\langle\sigma_p,e_p\rangle_{p \in N \setminus \{q\}}$ in
$\langle\gamma,s_0\rangle$.

There exist especially well-behaved Nash equilibria called
\emph{subgame perfect equilibria.} In order to define them,
we need the notion of an \emph{$n$-modification} of a strategy
$\sigma : E \to \bbC(E \times A)^B$.
Intuitively, an $n$-modification of~$\sigma$ is a new strategy
that coincides with~$\sigma$, except for the first $n$~turns of the game.
Formally, we define it as a strategy
\begin{align*}
  \sigma' : E + [n] \to \bbC((E + [n]) \times A)^B
\end{align*}
(where $[n] := \{0,\dots,n-1\}$) that satisfies the following conditions:
\begin{alignat*}{-1}
  \sigma'(e) &= \sigma(e)\,, &&\quad\text{for } e \in E\,, \\
  \sigma'(k) &\in \bbC(\{k+1\} \times A)^B &&\quad\text{for } k \in [n],\ k < n-1\,, \\
  \sigma'(k) &\in \bbC(E \times A)^B &&\quad\text{for } k = n-1\,.
\end{alignat*}

We say that a Nash equilibrium $\langle\bar\sigma,\bar e\rangle$ is
\emph{subgame perfect} if,
for every $n \in \bbN$, all $n$-modifications~$\bar\sigma'$ of~$\bar\sigma$,
and every player $q \in N$, $\langle\sigma'_q,0\rangle$
is a best response to
$\langle\sigma'_p,0\rangle_{p \in N \setminus \{q\}}$ in
$\langle\gamma,s_0\rangle$, where we restrict the notion of a best response
only to consider strategies coinciding with the given one in the first
$n$~turns.

\begin{example}
Consider the Repeated Prisoner's Dilemma introduced in
Example~\ref{ex: repeated prisoner's dilemma}.
To define the outcome of the game we use a discounted sum with
discount factor $\lambda < 1$.

We take a look at two strategies: (i)~a simple strategy~$\sigma$
that always denies
and (ii)~a `tit-for-tat' strategy~$\sigma'$
which mirrors the last move of the opponent.
We can define these two strategies as follows.
Both strategies use no epistemic states $E_p := \{*\}$
and as observations $B_p := \{c,d\}$ the last action of the opponent.
\begin{align*}
  \sigma(*)(x) := d
  \quad\text{and}\quad
  \sigma'(*)(x) := x\,.
\end{align*}

The pair $\langle\sigma,\sigma\rangle$ of simple strategies is a Nash
equilibrium with outcome $\langle0,0\rangle$, since every change of one
strategy results in a negative outcome for that player.
The equilibrium is subgame perfect, as both the strategies and the game
do not depend on the history of the play.

The pair $\langle\sigma',\sigma'\rangle$ of `tit-for-tat' strategies
is also a Nash equilibrium. Its outcome is
$\langle 2/(1-\lambda),2/(1-\lambda)\rangle$.
This time the equilibrium is not subgame perfect.
Consider the $1$-modification $\langle\sigma'_c,\sigma'_d\rangle$
of $\langle\sigma',\sigma'\rangle$ where in the first turn,
player~1 plays~$c$ while player~2 plays~$d$.
This leads to the play $(c,d)(d,c)(c,d)(d,c)\dots$ with outcome
$\frac{2-\lambda}{1-\lambda^2}$ (for the first player).
If, instead of~$\sigma'_d$, the second player chooses the strategy of always
playing~$d$, we obtain the play
$(c,d)(d,d)(d,d)(d,d)\dots$ with outcome~$2$. Since $0 < \lambda < 1$,
this is larger than $\frac{2-\lambda}{1-\lambda^2}$.
\end{example}

\subsection{Summary}   

Summing up the preceding sections, we have seen that we can specify a game
by the following data:
\begin{itemize}\itemsep=0pt\parskip=0pt%
\item a set~$N$ of \emph{players,}
\item for each $p \in N$, a set~$A_p$ of \emph{actions} for player~$p$,
\item for each $p \in N$, a set~$B_p$ of \emph{observations}
  for player~$p$,
\item a set~$S$ of \emph{states} of the game,
\item a set~$R$ of \emph{results,}
\item a set~$O$ of \emph{output values,}
\item a set~$U$ of \emph{outcomes,}
\item a function
  \begin{align*}
    \gamma : S  \to \bbC(R + S \times O)^{\prod_{p \in N} A_p}
  \end{align*}
  computing a single step of the game,
\item for each $p \in N$, a function
  \begin{align*}
    \beta_p : O \times \prod_{p \in N} A_p \to B_p
  \end{align*}
  computing the observations of player~$p$, and
\item two functions
  $\varrho : \Omega \to U$ and $\tau : \Gamma(U) \to U$ that satisfy
  \begin{align*}
    \varrho = \tau \circ \Gamma(\varrho) \circ \omega
  \end{align*}
  and thus compute the outcome of a play.
\end{itemize}


\section{Examples}\label{sec:examples}

In this section we present the formulations of two basic games in our framework.
The first game is one with imperfect information as imperfect monitoring.
The second game is the one with incomplete information \cite{shoham_multiagent_2008}.

The usual approach in game theory is to reduce incomplete information to imperfect one.
Incomplete information denotes situations where the type of agents is not known
while imperfect information denotes situations where the state of the game is not known.
In our framework this differentiation is not important since
both kinds of information deficiencies are captured by unobservable state spaces.
\subsection{Imperfect Public Monitoring}

The imperfect information game with imperfect monitoring and a noisy signal
considers games where the agents' actions may not be directly observable.
The state of the game is driven by a probabilistic state transition and
may be either ``good'' or ``bad''.
This information is publicly observed by the agents, i.e. all players
observe the same signal. The payoff is a function of this public outcome.

Again we have two players with two actions each: $N=\{1,2\}$
and $A_p=\{c, d\}$.
There are no results $R = \emptyset$ since the game never ends.
The output values are
$O=\bbR\times \bbR \times Y$ with $Y := \{G,B\}$ that
encode the payoffs in the stage games and the public signal.
The game has a single state $S = \{*\}$.

The game for the probabilistic functor $\bbC_\gamma = \bbC_\prob$,
is the function
\begin{align*}
    \gamma : S  &\to \bbC_\prob(S\times O)^{A_1\times A_2}\\
                    ({*}, a_1, a_2) &\mapsto ({*}, (r_1, r_2,y)) \\
\end{align*}
where
\begin{align*}
  r_p &= \begin{cases}
           1+\frac{2-2k}{k-m} &\text{if } (a_p,y)=(c, G) \\
           1-\frac{2k}{k-m}   &\text{if } (a_p,y)=(c, B) \\
           \frac{2-2n}{m-n}   &\text{if } (a_p,y)=(d, G) \\
           \frac{-2n}{m-n}    &\text{if } (a_p,y)=(d, B) \\
         \end{cases}\\
  y &= \begin{cases}
         G \text{ with probability } k   &\text{if } (a_1,a_2)=(c, c)   \\
         G \text{ with probability } m   &\text{if } (a_1,a_2)=(c, d)\lor (a_1,a_2)= (d,c)   \\
         G \text{ with probability } n   &\text{if } (a_1,a_2)=(d, d)   \\
         B \text{ with probability } 1-k &\text{if } (a_1,a_2)=(c, c)   \\
         B \text{ with probability } 1-m &\text{if } (a_1,a_2)=(c, d)\lor (a_1,a_2)= (d,c)   \\
         B \text{ with probability } 1-n &\text{if } (a_1,a_2)=(d, d)
       \end{cases}
\end{align*}
The probabilities of the state transition are characterized by the parameters
$k > m > n$. The parameters are chosen so that the expected value of the
payoffs is given by the prisoner's dilemma matrix:
\begin{align*}
\begin{array}{|c|c|c|}\hline
    & c      & d      \\ \hline
  c & (1,1)  & (-1,2) \\ \hline
  d & (2,-1) & (0,0)  \\ \hline
\end{array}
\end{align*}


The game has imperfect information so that the epistemic state of the players
is not the state of the game. Each player knows the history of his actions and
the history of the public signals
\begin{align*}
   E_p = (A_p \times Y)^*.
\end{align*}
The observation function is
\begin{align*}
  \beta_p : O \times A_1\times A_2 & \to B_p\\
              ((r_1,r_2,y), a_1, a_2) &\mapsto (r_p,y,a_p).
\end{align*}

We consider deterministic strategies with choice functor
$\bbC_\sigma = \bbC_\det$.
An example of always (unconditionally) playing~$d$ for a player~$p$ is given by
\begin{align*}
  \sigma_1 : E_1 &\to (E_1 \times A_1)^{B_1}\\
             (h,(r_1,y,a_1)) &\mapsto  (h(a_1,y),d)
\end{align*}
where $h(a_p,y)$ denotes that the history $h$ of the epistemic state is extended by the action $a_p$ and the public signal $y$.
\subsection{Incomplete Information}
In a Bayesian game of incomplete information types of agents are not common knowledge.
In the simplest case we take types of agents to be represented as different payoff functions of the game
and each agent knows his own type but not the one of his opponent.

In the following example we define that the game is played only once.
However, our framework is rich enough to easily extend the game to be played 
finitely, infinitely or potentially infinitely often.
The types of agents can be drawn repeatedly or as in the following example only once.
The agents can use Bayesian updating or in fact any kind of a learning rule. 

We define a Bayesian game of four $2\times 2$ games: 
Matching Pennies (MP), Prisoner's Dilemma (PD), Coordination Game (CG) and Battle of the Sexes (BS)
with equivalence classes $I_{i,j}$ for players $i$ and types $j$.
Player 1, if of type 1, knows that the payoff is either MP or PD and if of type 2, that the payoff is either CG or BS.
Player 2, if of type 1, knows that the payoff is either MP or CG and if of type 2, that the payoff is either PD or BS.
The probabilities are given by $p_{MP}=0.3, p_{PD}=0.1, p_{CG}=0.2$ and $p_{BS}=0.4$.

\begin{center}
\begin{tikzpicture}[scale=2.54]
\ifx\dpiclw\undefined\newdimen\dpiclw\fi
\global\def\dpicdraw{\draw[line width=\dpiclw]}
\global\def\dpicstop{;}
\dpiclw=0.8bp
\dpicdraw (0.535714,0.428571) rectangle (0.535714,0.428571)\dpicstop
\draw (0.535714,0.428571) node{MP};
\dpicdraw (0,0) rectangle (0.535714,0.357143)\dpicstop
\draw (0.267857,0.178571) node{2,0};
\dpicdraw (0.535714,0) rectangle (1.071429,0.357143)\dpicstop
\draw (0.803571,0.178571) node{0,2};
\dpicdraw (0,-0.357143) rectangle (0.535714,0)\dpicstop
\draw (0.267857,-0.178571) node{0,2};
\dpicdraw (0.535714,-0.357143) rectangle (1.071429,0)\dpicstop
\draw (0.803571,-0.178571) node{2,0};
\dpicdraw (0.535714,-0.428571) rectangle (0.535714,-0.428571)\dpicstop
\draw (0.535714,-0.428571) node{$p=0.3$};
\dpicdraw (1.964286,0.428571) rectangle (1.964286,0.428571)\dpicstop
\draw (1.964286,0.428571) node{PD};
\dpicdraw (1.428571,0) rectangle (1.964286,0.357143)\dpicstop
\draw (1.696429,0.178571) node{2,2};
\dpicdraw (1.964286,0) rectangle (2.5,0.357143)\dpicstop
\draw (2.232143,0.178571) node{0,3};
\dpicdraw (1.428571,-0.357143) rectangle (1.964286,0)\dpicstop
\draw (1.696429,-0.178571) node{3,0};
\dpicdraw (1.964286,-0.357143) rectangle (2.5,0)\dpicstop
\draw (2.232143,-0.178571) node{1,1};
\dpicdraw (1.964286,-0.428571) rectangle (1.964286,-0.428571)\dpicstop
\draw (1.964286,-0.428571) node{$p=0.1$};
\dpicdraw (0.535714,-0.642857) rectangle (0.535714,-0.642857)\dpicstop
\draw (0.535714,-0.642857) node{CG};
\dpicdraw (0,-1.071429) rectangle (0.535714,-0.714286)\dpicstop
\draw (0.267857,-0.892857) node{2,0};
\dpicdraw (0.535714,-1.071429) rectangle (1.071429,-0.714286)\dpicstop
\draw (0.803571,-0.892857) node{0,0};
\dpicdraw (0,-1.428571) rectangle (0.535714,-1.071429)\dpicstop
\draw (0.267857,-1.25) node{0,0};
\dpicdraw (0.535714,-1.428571) rectangle (1.071429,-1.071429)\dpicstop
\draw (0.803571,-1.25) node{1,1};
\dpicdraw (0.535714,-1.5) rectangle (0.535714,-1.5)\dpicstop
\draw (0.535714,-1.5) node{$p=0.2$};
\dpicdraw (1.964286,-0.642857) rectangle (1.964286,-0.642857)\dpicstop
\draw (1.964286,-0.642857) node{BS};
\dpicdraw (1.428571,-1.071429) rectangle (1.964286,-0.714286)\dpicstop
\draw (1.696429,-0.892857) node{2,1};
\dpicdraw (1.964286,-1.071429) rectangle (2.5,-0.714286)\dpicstop
\draw (2.232143,-0.892857) node{0,0};
\dpicdraw (1.428571,-1.428571) rectangle (1.964286,-1.071429)\dpicstop
\draw (1.696429,-1.25) node{0,0};
\dpicdraw (1.964286,-1.428571) rectangle (2.5,-1.071429)\dpicstop
\draw (2.232143,-1.25) node{1,2};
\dpicdraw (1.964286,-1.5) rectangle (1.964286,-1.5)\dpicstop
\draw (1.964286,-1.5) node{$p=0.4$};
\dpicdraw (-0.142857,-0.517857) rectangle (2.714286,0.517857)\dpicstop
\dpicdraw (-0.142857,-1.589286) rectangle (2.714286,-0.553571)\dpicstop
\dpicdraw[dashed](-0.071429,-1.642857) rectangle (1.142857,0.571429)\dpicstop
\dpicdraw[dashed](1.357143,-1.642857) rectangle (2.571429,0.571429)\dpicstop
\dpicdraw (-0.285714,0) rectangle (-0.285714,0)\dpicstop
\draw (-0.285714,0) node{$I_{1,1}$};
\dpicdraw (-0.285714,-1.071429) rectangle (-0.285714,-1.071429)\dpicstop
\draw (-0.285714,-1.071429) node{$I_{1,2}$};
\dpicdraw (0.535714,0.642857) rectangle (0.535714,0.642857)\dpicstop
\draw (0.535714,0.642857) node{$I_{2,1}$};
\dpicdraw (1.964286,0.642857) rectangle (1.964286,0.642857)\dpicstop
\draw (1.964286,0.642857) node{$I_{2,2}$};
\end{tikzpicture}

\end{center}

The state space of the game is $S=\{*,MP, PD, CG, BS\}$.
The output space of the game is $O=S$ and the result space is $R=\bbR \times \bbR$.
The action spaces are $A_1=\{U,D\}$ and $A_2=\{L,R\}$ and the epistemic state spaces have a single state $E_1=E_2=\{*\}$.
We formalize this game in two rounds.

\begin{enumerate}
\item In the first round the game is in state $*$ and nature realizes the types of the players, the actions of the players are irrelevant.
\begin{align*}
    \gamma :  S &\to \bbC_\prob(R + S \times O)^{A_1 \times A_2}\\
                    {(*,a_1,a_2)} &\mapsto 
  \begin{cases}
         (MP,MP) &\text{ with } p_{MP}=0.3  \\
         (PD,PD) &\text{ with } p_{PD}=0.1  \\
         (CG,CG) &\text{ with } p_{CG}=0.2  \\
         (BS,BS) &\text{ with } p_{BS}=0.4  \\
  \end{cases} 
\end{align*} 
In the second round the game yields a result.
\begin{align*}
                      (MP, a_1, a_2) &\mapsto 
  \begin{cases}
         (2,0) &\text{ if } a_1=U, a_2=L  \\
         (0,2) &\text{ if } a_1=U, a_2=R  \\
         (0,2) &\text{ if } a_1=D, a_2=L  \\
         (2,0) &\text{ if } a_1=D, a_2=R  \\
  \end{cases}
\end{align*}
An analogous definition has to be given for the other type realizations $PD, CG, BS$.
\item The observation function of the first player is
\begin{align*}
  \beta_1 : O \times A_1 \times A_2 &\to B_1\\
                    {(o,a_1,a_2)} &\mapsto 
  \begin{cases}
         * &\text{ if } o=*  \\
         \{MP,PD\} &\text{ if } o=MP \lor o=PD  \\
         \{CG,BS\} &\text{ if } o=CG \lor o=BS  \\
  \end{cases}
\end{align*}
  The observation function of player 2 is defined analogously.
\item The strategies for both players are given by
\begin{align*}
  \sigma_p : \{*\} &\to \bbC(\{*\}\times A_p)^{B_p} \\
\end{align*}
For example, the strategy of player 1 who plays $U$ if he is of type 1 and $D$ if he is of type 2 is given by
\begin{align*}
                    (*, b_1) &\mapsto 
  \begin{cases}
         U \text{ if } b_1=\{MP,PD\} \\
         D \text{ if } b_1=\{CG,BS\} \\
  \end{cases}
\end{align*}
In the first round when nature chooses the types, the strategy of the players is irrelevant since it does not matter in the game function.
\end{enumerate}
%



\bibliographystyle{plain}
\bibliography{library}

\begin{thebibliography}{10}

\bibitem{abramsky_introduction_2011}
S.~Abramsky and N.~Tzevelekos.
\newblock Introduction to categories and categorical logic.
\newblock In Bob Coecke, editor, {\em New Structures for Physics}. Springer,
  Berlin, 2011.

\bibitem{abramsky_lawvere_2010}
S.~Abramsky and J.~Zvesper.
\newblock From {L}awvere to {B}randenburger-{K}eisler: {I}nteractive {F}orms of
  {D}iagonalization and {S}elf-reference.
\newblock In Dirk Pattinson and Lutz Schr{\"o}der, editors, {\em CMCS}, number
  7399 in Lecture Notes in Computer Science, pages 1--19. Springer, 2012.

\bibitem{abramsky_coalgebraic_2012}
Samson Abramsky and Viktor Winschel.
\newblock Coalgebraic analysis of subgame-perfect equilibria in infinite games
  without discounting.
\newblock {\em {arXiv:1210.4537}}, 2012.

\bibitem{harsanyi_games_1967}
John~C. Harsanyi.
\newblock Games with incomplete information played by {"Bayesian"} players,
  i-{III.} part i. the basic model.
\newblock {\em Management Science}, 14(3):159--182, 1967.

\bibitem{harsanyi_games_1968}
John~C. Harsanyi.
\newblock Games with incomplete information played by {"Bayesian"} players,
  i-{III.} part {II.} bayesian equilibrium points.
\newblock {\em Management Science}, 14(5):320--334, 1968.

\bibitem{harsanyi_games_1968-1}
John~C. Harsanyi.
\newblock Games with incomplete information played by {"Bayesian"} players,
  i-{III.} part {III.} the basic probability distribution of the game.
\newblock {\em Management Science}, 14(7):486--502, 1968.

\bibitem{heifetz_topology-free_1998}
Aviad Heifetz and Dov Samet.
\newblock Topology-free typology of beliefs.
\newblock {\em Journal of Economic Theory}, 82(2):324--341, 1998.

\bibitem{honsell_conway_2011}
F.~Honsell, M.~Lenisa, and R.~Redamalla.
\newblock Conway games, algebraically and coalgebraically.
\newblock {\em Logical Methods in Computer Science}, 7(3:08):1--30, 2011.

\bibitem{jacobs_introduction_2012}
Bart Jacobs.
\newblock {\em Introduction to Coalgebra. Towards Mathematics of States and
  Observations}.
\newblock 2012.

\bibitem{klin_bialgebras_2011}
B.~Klin.
\newblock Bialgebras for structural operational semantics: An introduction.
\newblock {\em Theoretical Computer Science}, 412(38):5043--5069, 2011.

\bibitem{kurz_coalgebras_2006}
A.~Kurz.
\newblock Coalgebras and their logics.
\newblock {\em {ACM} {SIGACT} News}, 37(2):57--77, 2006.

\bibitem{lawvere_diagonal_1969}
F.~William Lawvere.
\newblock Diagonal arguments and cartesian closed categories.
\newblock In {\em Category Theory, Homology Theory and their Applications
  {II}}, volume~92, pages 134--145. Springer Berlin Heidelberg, 1969.

\bibitem{lescanne_backward_2012}
Pierre Lescanne and Matthieu Perrinel.
\newblock {``Backward''} coinduction, nash equilibrium and the rationality of
  escalation.
\newblock {\em Acta Informatica}, 49(3):117--137, 2012.

\bibitem{luhmann_gesellschaft_1998}
Niklas Luhmann.
\newblock {\em Die Gesellschaft der Gesellschaft}.
\newblock Suhrkamp Verlag, 7th edition, 1998.

\bibitem{mac_lane_categories_1998}
Saunders Mac~Lane.
\newblock {\em Categories for the Working Mathematician}.
\newblock Springer, 2nd edition, 1998.

\bibitem{mertens_formulation_1985}
Jean-Fran{\c c}ois Mertens and Shmuel Zamir.
\newblock Formulation of bayesian analysis for games with incomplete
  information.
\newblock {\em International Journal of Game Theory}, 14(1):1--29, 1985.

\bibitem{moss_final_2006}
L~Moss and I~Viglizzo.
\newblock Final coalgebras for functors on measurable spaces.
\newblock {\em Information and Computation}, 204(4):610--636, 2006.

\bibitem{moss_harsanyi_2004}
Lawrence~S. Moss and Ignacio~D. Viglizzo.
\newblock Harsanyi type spaces and final coalgebras constructed from satisfied
  theories.
\newblock {\em Electronic Notes in Theoretical Computer Science}, 106:279--295,
  2004.

\bibitem{pavlovic_semantical_2009}
Dusko Pavlovic.
\newblock A semantical approach to equilibria and rationality.
\newblock In Alexander Kurz, Marina Lenisa, and Andrzej Tarlecki, editors, {\em
  Algebra and Coalgebra in Computer Science}, number 5728 in Lecture Notes in
  Computer Science, pages 317--334. Springer Berlin Heidelberg, 2009.

\bibitem{plotkin_structural_2004}
Gordon Plotkin.
\newblock A structural approach to operational semantics.
\newblock {\em The Journal of Logic and Algebraic Programming},
  60--61(0):17--139, 2004.

\bibitem{rutten_universal_2000}
J.~J. M.~M. Rutten.
\newblock Universal coalgebra: a theory of systems.
\newblock {\em Theoretical Computer Science}, 249(1):3--80, 2000.

\bibitem{schmidt_denotational_1986}
D.~A Schmidt.
\newblock {\em Denotational Semantics: A Methodology for Language Development}.
\newblock Allyn and Bacon, Inc., 1986.

\bibitem{shoham_multiagent_2008}
Yoav Shoham and Kevin Leyton-Brown.
\newblock {\em Multiagent Systems: Algorithmic, Game-Theoretic, and Logical
  Foundations}.
\newblock Cambridge University Press, 2008.

\bibitem{trancon_y_widemann_distributive-law_2011}
Baltasar Trancon~y Widemann and Michael Hauhs.
\newblock Distributive-law semantics for cellular automata and agent-based
  models.
\newblock In {\em Proceedings of the 4th international conference on Algebra
  and coalgebra in computer science}, pages 344--358. Springer Berlin
  Heidelberg, 2011.

\bibitem{yanofsky_universal_2003}
N.~S Yanofsky.
\newblock A universal approach to self-referential paradoxes, incompleteness
  and fixed points.
\newblock {\em Bulletin of Symbolic Logic}, 9(3):362--386, 2003.

\end{thebibliography}
\end{document}